\documentclass[aps,prmaterials,twocolumn,superscriptaddress]{revtex4-2}
\usepackage[utf8]{inputenc}
\usepackage{longtable}
\usepackage{amsmath}%
\usepackage{amsfonts}%
\usepackage{amssymb}%
\usepackage{graphicx}
\usepackage[colorlinks=true,linkcolor=blue,urlcolor=black,hyperfootnotes=true,citecolor=blue]{hyperref}
\usepackage{CJKutf8} 
\usepackage{ulem} 
\usepackage[product-units=power, separate-uncertainty=true, multi-part-units=single]{siunitx} 
\usepackage[usenames,dvipsnames]{xcolor}
\usepackage{threeparttable}

\DeclareMathOperator{\sgn}{sgn}
\newcommand{\fig}{Fig.~\ref}
\newcommand{\figs}{Figs.~\ref}
\newcommand{\dIdU}{$\dd I / \dd U$}
\newcommand{\dd}{\text{d}}
\newcommand{\SrBiSe}{${\mathrm{Sr}}_x{\mathrm{Bi}}_{2}{\mathrm{Se}}_{3}$}  
\newcommand{\BiSe}{${\mathrm{Bi}}_{2}{\mathrm{Se}}_{3}$}
\newcommand{\scanp}{Scan parameter:}
\newcommand{\stabp}{Stabilization parameter:}
\newcommand{\diduiu}{($\dd I / \dd U$)$/$($I/U$)}

\newcommand{\myref}[7]{\href{http://dx.doi.org/#7}{#1, #2, \color{blue}#3 \textbf{#4}, #5 (#6).}} 
\newcommand{\textref}[1]{{#1}} 

\newcommand{\etal}{\textit{et al.}}

\begin{document}

\title{Observability of superconductivity in Sr-doped Bi$_2$Se$_3$ at the surface 
	  using scanning tunneling microscope}

\author{Mahasweta Bagchi}     \affiliation{Physics Institute II, University of Cologne, D-50937 Köln, Germany}

\author{Jens Brede}  \email{brede@ph2.uni-koeln.de}    \affiliation{Physics Institute II, University of Cologne, D-50937 Köln, Germany}

\author{Yoichi Ando }   \email{ando@ph2.uni-koeln.de}    \affiliation{Physics Institute II, University of Cologne, D-50937 Köln, Germany}

\date{\today}

\begin{abstract}
The superconducting materials family of doped ${\mathrm{Bi}}_{2}{\mathrm{Se}}_{3}$ remains intensively studied in the field of condensed matter physics due to strong experimental evidence for topologically non-trivial superconductivity in the bulk. However, at the surface of these materials, even the observation of superconductivity itself is still controversial.
We use scanning tunneling microscopy (STM) down to 0.4~K to show that on the surface of bulk superconducting ${\mathrm{Sr}}_x{\mathrm{Bi}}_{2}{\mathrm{Se}}_{3}$, no gap in the density of states is observed around the Fermi energy as long as clean metallic probe tips are used. Nevertheless, using scanning electron microscopy and energy-dispersive X-ray analysis, we find that micron-sized flakes of ${\mathrm{Sr}}_x{\mathrm{Bi}}_{2}{\mathrm{Se}}_{3}$ are easily transferred from the sample onto the STM probe tip and that such flakes consistently show a superconducting gap in the density of states. 
We argue that the superconductivity in ${\mathrm{Sr}}_x{\mathrm{Bi}}_{2}{\mathrm{Se}}_{3}$ crystals does not extend to the surface when the topological surface state (TSS) is intact, but in micro-flakes the TSS has been destroyed due to strain and allows the superconductivity to extend to the surface. To understand this phenomenon, we propose that the local electric field, always found in electron doped ${\mathrm{Bi}}_{2}{\mathrm{Se}}_{3}$ in the presence of the TSS due to an intrinsic upward band bending, works against superconductivity at the surface.

\end{abstract}

\maketitle

\section{Introduction}

Shortly after the discovery of superconductivity in Cu-doped \BiSe{} crystals~\cite{Hor2010}, Fu and Berg~\cite{Fu2010} proposed that any electron-doped \BiSe{} is a viable candidate for hosting topological superconductivity with spin-triplet-like pairing. The spin-triplet-like nature of the pairing was successively confirmed by temperature-dependent nuclear magnetic resonance Knight shift ($K_\mathrm{s}$) experiments~\cite{Matano2016}, which found no change in $K_\mathrm{s}$ below $T_\mathrm{c}$ for magnetic fields applied parallel to the $c$-axis. Moreover, the same experiments found that the three-fold-symmetric \BiSe{} lattice showed a two-fold anisotropy of $K_\mathrm{s}$ when the magnetic field was rotated in the $ab$ plane. This indicates a spontaneous rotational symmetry breaking of the superconducting state. The two-fold symmetry of the superconducting state was also observed in specific heat~\cite{Yonezawa2017}, which indicates that this symmetry breaking is due to an anisotropy in the superconducting gap amplitude and points to nematic superconductivity \cite{Yonezawa2019}. A recent high-resolution x-ray diffraction (XRD) experiment clarified~\cite{Kuntsevich2018} that a tiny ($\sim$0.02\%) lattice distortion dictates the nematic axis. Theoretically, this nematic superconductivity is expected in doped \BiSe{} superconductors for the superconducting gap function having $E_u$ symmetry \cite{Fu2010, Ando2015, Sato2017}, which is topologically non-trivial.

Concurrently with the bulk characterization, surface sensitive techniques, in particular scanning tunneling microscopy (STM) and spectroscopy (STS), were used to study the superconducting properties at the surface of doped \BiSe{} crystals. 
Already in 2013 Levy \etal~\cite{Levy2013} reported the observation of both normal-conducting and superconducting (SC) domains at the surface of Cu-doped \BiSe{}. These SC domains showed a fully gapped local density of states (LDOS) at the Fermi level that could be well-described within the Bardeen–Cooper–Schrieffer (BCS) theory. The observed gap width was $\Delta=0.4$~meV at the surface and bulk resistance vs.\ temperature measurements showed a superconducting transition at around $3.65$~K. 
Moreover, vortices with a diameter of about 30~nm were observed at the surface under an applied out-of-plane magnetic field of more than 0.5~T, and the upper critical field was determined to be $\mu_0 H_\mathrm{c2}\approx 1.65$~T.
Interestingly, no zero bias conductance peaks were observed in the vortex core by Levy \etal~\cite{Levy2013} while a more recent STM study of Cu-doped \BiSe{} by Tao \etal~\cite{Tao2018} resolved an Abrikosov lattice consisting of elliptically-shaped vortices on the surface, which also hosted a zero bias conductance peak. 
However, Tao \etal~\cite{Tao2018} also documented two different SC domains with largely different gap sizes of 0.46~meV and 0.77~meV, and 96\% of the surface areas they studied did not show any superconductivity.

Such differences in the superconducting properties observed at the surface of Cu-doped \BiSe{} may be related to the comparatively poor superconducting volume fraction of only about $40-50\%$~\cite{Kriener2011} and associated inhomogeneity of the superconducting phase throughout the sample.
In this regard, \SrBiSe{} is better suited to STM studies since the superconducting volume fraction reaches more than $90\%$~\cite{Liu2015} and thus one expects to avoid the ambiguity between local probe and bulk measurements that arises due to inhomogeneity.

However, at the surface of a \SrBiSe{} crystal with a bulk $T_\mathrm{c}$ of 2.4~K, Han \etal~\cite{Han2015} observed a superconducting gap which only dropped to $75\%$ of the normal state conductance at the Fermi energy despite a gap size of $\Delta\approx0.5$~meV. The authors attributed their observation to their relatively high measurement temperature of 1~K.
In 2017, Du {\it et al.} \cite{Du2017} observed on the surface of a \SrBiSe{} crystal with a bulk $T_\mathrm{c}$ of 3 K, SC domains with the gap size of $\Delta\approx$ 0.42 -- 1.15 meV. Moreover, for a domain with $\Delta\approx 0.8$~meV a $T_\mathrm{c}$ = 5~K and $\mu_0 H_\mathrm{c2}\approx 5$~T was reported. Note that in the bulk of \SrBiSe{}, $\mu_0 H_\mathrm{c2}$ amounts only to about 1.5~T  \cite{Shruti2015}. 
A recent work by Kumar \etal~\cite{Kumar2021} reported a superconducting gap of $\Delta\approx 0.19$ -- 0.31~meV on the surface of a \SrBiSe{} crystal with the bulk $T_\mathrm{c}$ of 2.9~K.
  
With regards to such inconsistencies, Wilfert \etal~\cite{Wilfert2018} recently showed that for Tl-doped Bi$_2$Te$_3$ and Nb-doped \BiSe{}, the superconducting gaps on the surface were exclusively observed due to the nominally normal conducting probe tips becoming unintentionally superconducting during the experiments. Similar experimental pitfalls were also reported for Cu-doped \BiSe{} by Levy \etal~\cite{Levy2013}.
Interestingly, Wilfert \etal~\cite{Wilfert2018} concluded that for Tl-doped Bi$_2$Te$_3$ and Nb-doped \BiSe{}, superconductivity does not extend to the surface where the topological surface state resides. 

Here, to clarify this complicated situation, we use transport measurements to characterize the bulk $T_\mathrm{c}$ and the carrier concentration of high-quality single crystals of \SrBiSe{} and subsequently perform high-resolution studies of STM and STS on the surface of crystals which are cleaved under ultra high vacuum conditions (UHV). Our base temperature is $0.4$~K.

\section{Experimental Methods}

\textit{Crystal growth.} Single crystals of \SrBiSe{} (nominal $x = 0.06$) are grown from high-purity elemental Sr chunk (99.99\%), Bi shots (99.9999\%), and Se shots (99.9999\%) by a conventional melt-growth method. The raw materials with a total weight of $4.0\,\mathrm{g}$ are mixed and sealed in an evacuated quartz tube. The tube is heated to $850^\circ\mathrm{C}$ for $48\, \mathrm{h}$. It is then slowly cooled from $850^\circ\mathrm{C}$ to $600^\circ\mathrm{C}$ within $80\,\mathrm{h}$ and finally quenched into water at room temperature.

\textit{Transport measurements.} Resistivity and Hall measurements on the samples are performed in a Quantum Design Physical Properties Measurement System (PPMS) in the standard four-terminal configuration using a low-frequency ac lock-in technique.

\textit{STM measurements.} STM experiments are carried out under UHV conditions with a commercial system (Unisoku USM1300) operating at $0.4$~K. Data are acquired at $0.4$~K unless mentioned otherwise. Topograph and \dIdU{} maps are recorded in the constant-current mode. Point spectroscopy data is obtained by first stabilizing for a given set-point condition and then disabling the feedback loop. \dIdU{} curves are then recorded by means of a lock-in amplifier by adding a small modulation voltage $U_{\text{mod}}$ to the sample bias voltage $U$. High resolution \dIdU{} spectra of superconducting gaps were normalized by fitting a second degree polynomial to the data outside the SC gap and dividing by the fitted polynomial.
We have used both PtIr and W probe tips. All PtIr tips used are commercially obtained from Unisoku. The W tips are made in-house. Both types are electrochemically etched. The PtIr tips are either fresh new tips or they have been prepared by Ar ion sputtering (at an argon pressure of $3\times10^{-6}$~mbar and a voltage of $1$~kV), followed by repeated heating by electron bombardment ($\sim15$~W) for $20$~s. Further tip forming is done by scanning on the Cu(111) surface until a clean signature of the surface state is obtained in spectroscopy. The absence of a superconducting gap on Cu(111) is also verified prior to measurements on \SrBiSe{}. For STM measurements, \SrBiSe{} crystals are cleaved at room temperature and under UHV conditions. The crystal is cleaved by breaking off a 10~mm sized pole glued on the sample. The two-component epoxy glue (EPO-TEK H21D) is hardened by heating to $373$~K under high vacuum conditions. STM data are processed using the WSxM software \cite{wsxm} and Igor Pro 9.0.

\textit{SEM and EDX analysis.} Scanning electron microscope (SEM) image of the tip is obtained using the Raith Pioneer II system and the Jeol JSM-6510 SEM. Elemental chemical analysis of the material on the tip apex is done by energy-dispersive X-ray (EDX) analysis, performed using an Oxford Instruments AztecOne system with a x-act Silicon Drift Detector that is combined with the Jeol SEM.

\section{Results and discussion}
\label{sec:Results-STM}

\subsection{Bulk properties}

\begin{figure}[t!]
	\includegraphics[width=8.6cm]{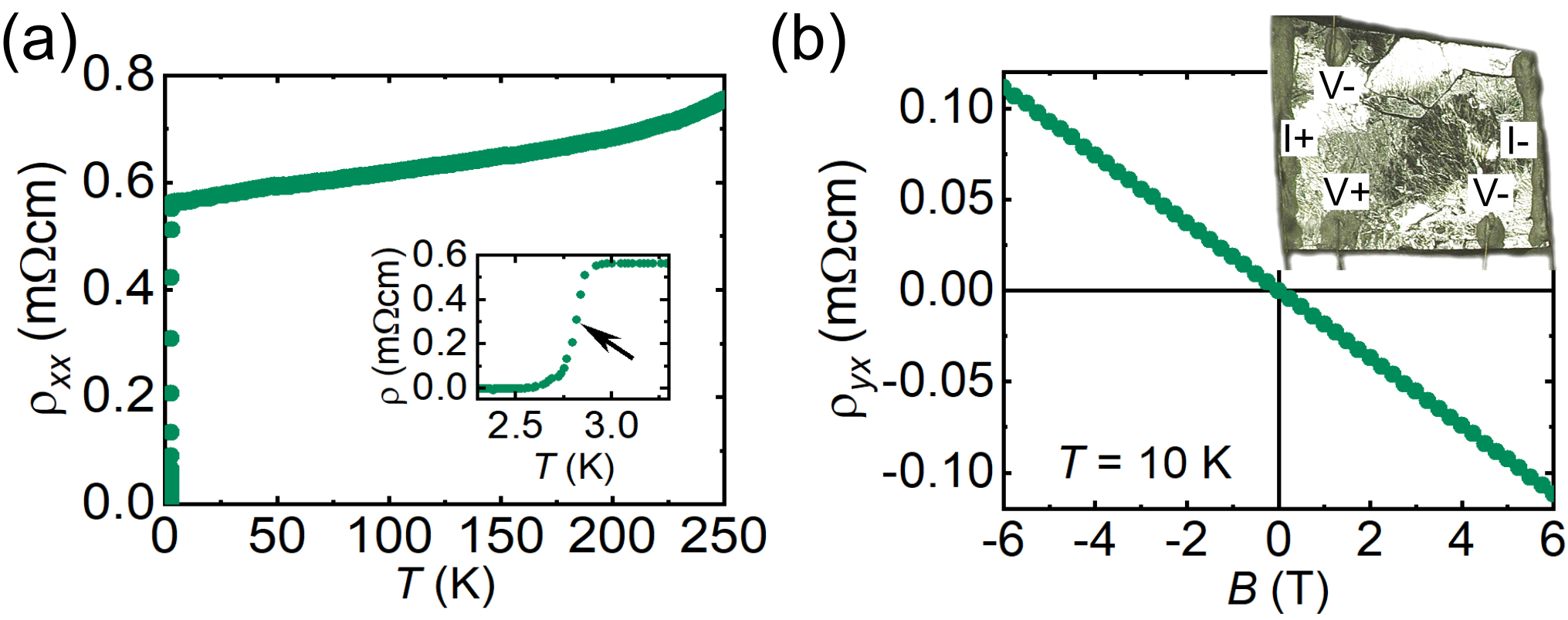}
	\caption{(a) Temperature dependence of the resistivity ($\rho_{xx}$) of a \SrBiSe{} crystal used for this work. The inset shows the superconducting transition with the mid-point $T_\mathrm{c}$ = 2.8~K. (b) Magnetic-field dependence of the Hall resistivity ($\rho_{yx}$) measured at $10$~K, which gives the carrier density of $3.4\times10^{19}$~$\text{cm}^{-3}$. The inset shows a picture of the \SrBiSe{} sample with contacts for resistivity and Hall measurements.
	}
	\label{fig:STM_hall}
\end{figure}

We have characterized the bulk properties of our \SrBiSe{} single crystals using resistivity ($\rho_{xx}$) and Hall resistivity ($\rho_{yx}$) measurements. An optical image of a typical sample ($5~\mathrm{mm} \times 4
~\mathrm{mm} \times 0.7~\mathrm{mm}$) including the electrical contacts is shown in the inset of \fig{fig:STM_hall}(b). The temperature dependence of the resistivity is metallic [\fig{fig:STM_hall}(a)] with the onset of superconductivity and zero-resistivity occurring at $2.90$ and $2.65$~K, respectively [inset of \fig{fig:STM_hall}(a)]. The superconducting transition temperature $T_\mathrm{c}$, defined by the mid-point of the resistive transition, is 2.8~K. The residual resistivity of 0.56~m$\Omega$cm, which points to a high scattering rate that is always present in doped \BiSe{} superconductors, is unusually large for an unconventional non-$s$-wave superconductor; nevertheless, it has been elucidated that in doped \BiSe{} superconductors where the orbital degrees of freedom play an important role, the generalized Anderson's theorem protects the unconventional pairing from disorder \cite{Andersen2018,Andersen2020}.

The carrier density in our samples is determined from $\rho_{yx}$ measured at $10$~K as a function of perpendicular magnetic field $B$ [\fig{fig:STM_hall}(b)]. The $\rho_{yx}(B)$ behavior is strictly linear in $B$ and can be described by a single band, yielding a carrier density of $n=3.4\times10^{19}$~$\text{cm}^{-3}$, which is extremely low for a superconductor with $T_\mathrm{c}$ of the order a few Kelvin.
Table \ref{tab:carrier_density} gives an overview of the carrier density of \SrBiSe{} samples (varying nominal doping) as reported in literature along with the values observed for our samples.
From Shubnikov-de Haas investigations by K\"ohler \etal~\cite{Koehler1973} it is known that a carrier density of $n\approx4\times10^{19}$~$\text{cm}^{-3}$ in \BiSe{} corresponds to a Fermi energy of $E_\mathrm{F}\approx160$~meV, which we use as a lower bound for the Fermi energy in the bulk of our \SrBiSe{} crystals. As an upper bound one can assume a simple parabolic dispersion for the bulk conduction band (BCB). Here, a carrier density of $n\approx4\times10^{19}$~$\text{cm}^{-3}$ corresponds to $E_\mathrm{F}\approx270$~meV for an effective mass of $m_\mathrm{eff}=0.15m_e$ \cite{Koehler1973,Analytis2010}, with $m_e$ the free electron mass.

The shielding fraction of our samples estimated from the zero-field-cooled magnetization measurement lies between 75\% to 100\% \cite{Lin2021}. The actual data of one of the samples measured here, which showed the shielding fraction of 76\%, were previously shown in Ref.~\cite{Lin2021}.

\subsection{Surface properties}
\label{sec:surface}

\begingroup
\setlength{\tabcolsep}{8pt}
\renewcommand{\arraystretch}{1.2}
\begin{table}
	\begin{tabular}{c c c}
		\hline\hline
		
		Reference  &  nominal Sr doping & carrier density  \\
		& & $n$ ($10^{19}$~$\text{cm}^{-3}$) \\
		\hline
		Liu 2015 \cite{Liu2015} &  0.062 & 2.65  \\
		Shruti 2015 \cite{Shruti2015} &  0.1 & 1.85\\
		Huang 2017 \cite{Huang2017} &  0.066 & 2.75\\
		Kuntsevich 2019 \cite{Kuntsevich2019} & 0.064, 0.068 & 2.2, 2.1\\
		Li 2018 \cite{Li2018} &  0.05 & 5.7--10\\
		Li 2018 \cite{Li2018} &  0.08 & 6.8--9.2\\
		this work &  0.06 & 3.4--6.2\\
		
		\hline\hline
	\end{tabular}
	\caption{Summary of bulk carrier density $n$ reported for \SrBiSe{}.}
	\label{tab:carrier_density}
\end{table}
\endgroup

\begin{figure}[t!]
	\includegraphics[width=8.6cm]{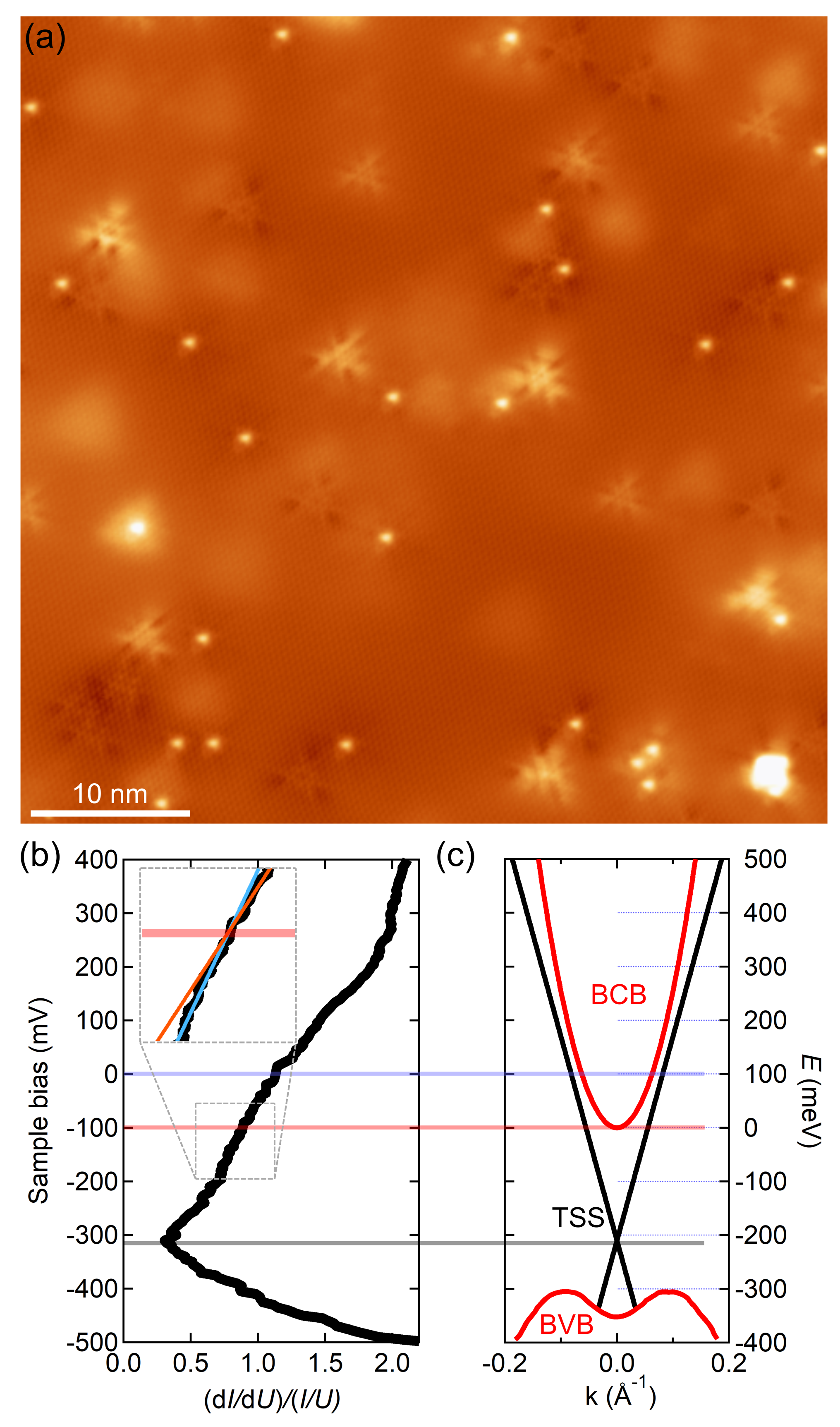}
	\caption{(a) Typical STM image of the (001) surface of \SrBiSe{} crystal directly after cleaving. Only a small number of native defects (discussed in Ref. \cite{Lin2021}) are visible. \scanp{} $U=+100$~mV, $I=20$~nA. (b) Representative \diduiu{} spectrum taken far away from any defect; inset shows a magnification of the range indicated by the gray dashed box to highlight a slight change in slope across $-100$ mV (red and blue lines with different slopes are a guide to the eye). \stabp{} $U=-900$~mV, $I=2$~nA, $U_{\text{mod}}= 10$~mV$_{\text{p}}$. (c) Schematic diagram of the band structure of \SrBiSe{}: zero energy is set at the bottom of the bulk conduction band (BCB). Grey, red and blue lines from (b) to (c) mark the Dirac point (DP) of the topological surface state (TSS), bottom of the conduction band, and the Fermi level, respectively. 
	The Fermi level at the surface lies at $\sim$100~meV in the BCB.
	}
	\label{fig:STM_STS}
\end{figure}

A typical topograph of the cleaved \SrBiSe{} surface is shown in \fig{fig:STM_STS}(a). The surface is atomically flat with some characteristic native defects, which we have previously discussed in detail \cite{Lin2021}. A representative \diduiu{} spectrum, which is proportional to the the local density of states (LDOS), is shown in \fig{fig:STM_STS}(b). The minimum of the LDOS is at $-310$~mV and corresponds to the Dirac point (DP).  
Based on the band structure of \BiSe{}, which is well-known from ARPES experiments \cite{Han2015,Neupane2016,Xia2009,Analytis2010} and schematically depicted in \fig{fig:STM_STS}(c), we assign the increase in slope at around $-400$~mV and below to the onset of the bulk valence band (BVB), and the increase at $-100$~mV and above to the BCB, respectively.
Therefore, at the \textit{surface} of this sample, the Fermi energy lies about $100$~meV above the bottom of the BCB, which is much lower than the estimate of $E_{\rm F} \approx 160$ -- 270~meV based on transport measurements.
However, this apparent disagreement is straightforwardly reconciled if we consider band bending to be present at the surface of electron-doped \BiSe{}. 

Band bending occurs due to charge transfer caused by the equilibration of the Fermi level at an interface. The charge transfer creates an electric field and the associated potential shifts the bands in the vicinity. 
In the case of a topological insulator, the existence of the TSS causes the charge distribution to be different near the surface compared to the bulk. When the TSS is electron-doped, the electrons in the TSS can be viewed as a negative surface charge $\sigma_s$. This surface charge is related to the surface potential $V_0=V(z=0)\propto -\sigma_s$ through Poisson's equation and the condition of overall charge neutrality. Hence, the surface charge in the TSS causes a positive potential leading to upward band bending of the BCB when going from the bulk to the surface. In other words, at the surface, charge equilibration causes fewer electrons in the BCB than in the bulk. For highly doped semiconductors, the decay of the potential into the bulk may be estimated within the Thomas-Fermi screening model as $V(z)=V_0\exp{(-z/r_\mathrm{TF})}$, where $r_\mathrm{TF}\approx \sqrt{(\epsilon_0\pi^2\hbar^2)/(k_\mathrm{F}m_\mathrm{eff}e^2)}\approx0.6$~nm is the Thomas-Fermi screening length and $k_\mathrm{F}\approx0.7$~\AA$^{-1}$. 
Interestingly, calculations within the density functional theory (DFT) in Refs.~\cite{Fregoso2015,Rakyta2015} show that even for pristine \BiSe{} an intrinsic upward band bending of the BCB of the order of $\sim100$~meV takes place due to charge equilibration between bulk-like states and the TSS when the Fermi energy lies above the DP. In these calculations, the BCB has recovered its bulk value at $2$ or $3$~nm below the surface.

\begin{figure}[h!]
	\includegraphics[width=8.6cm]{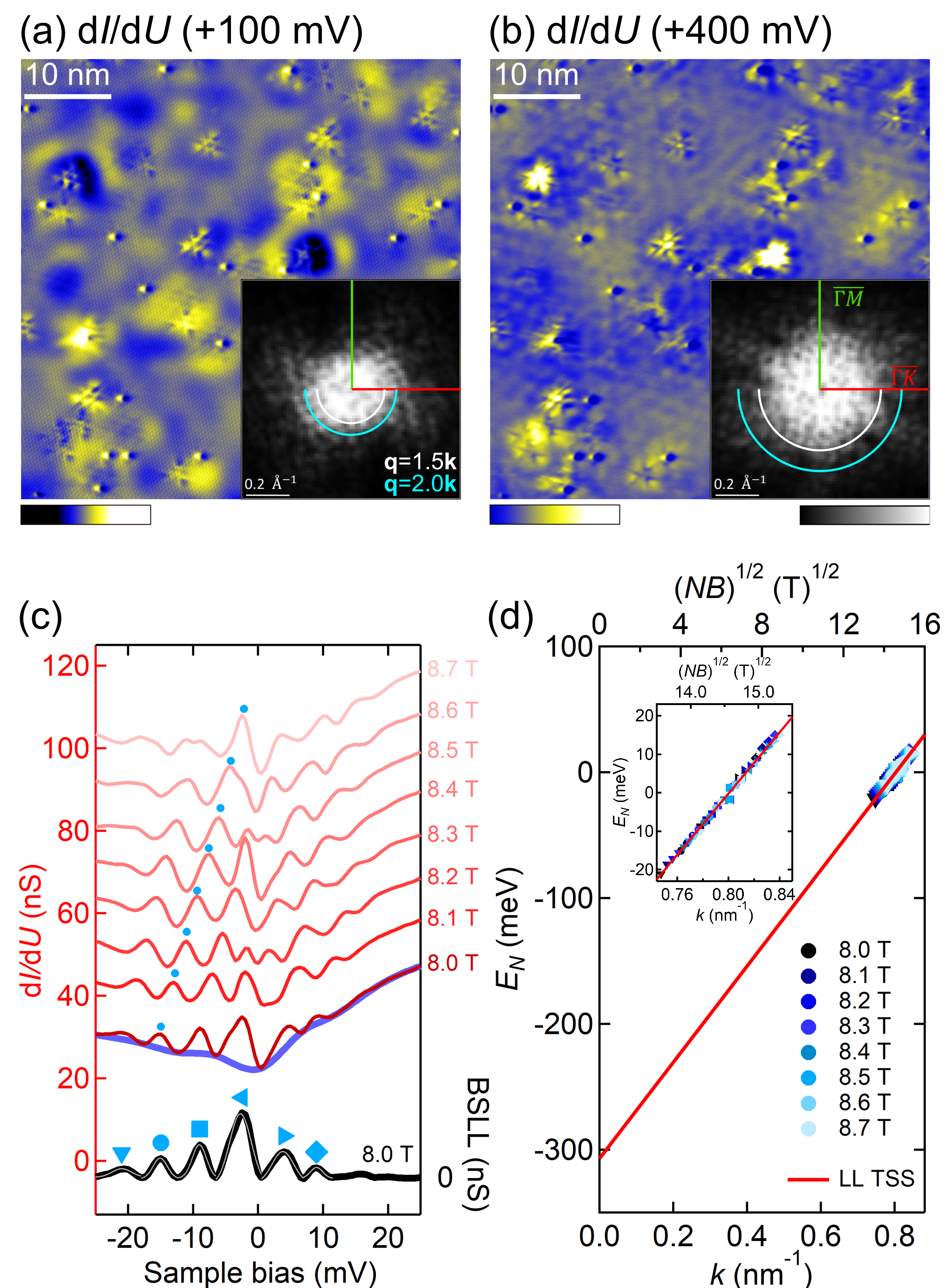}
	\caption{(a,b) Differential conductance images at the sample bias of $+100$~mV (a) and $+400$~mV (b); the insets provide the corresponding Fourier-transform images. Blue and white semi-circles mark the expected positions of the scattering vector \textbf{q} with \textbf{q} = 2\textbf{k} and 1.5\textbf{k}, respectively, with \textbf{k} the wavevector of the TSS. Set-point current: 20~nA. (c) \dIdU{} spectra recorded in the perpendicular magnetic field from $8$ to $8.7$~T with $0.1$~T interval. Spectra are shifted vertically for clarity. All spectra are taken with stabilization parameters of 	$U=50$~mV, $I=2$~nA, $U_{\text{mod}}= 1$~mV$_{\text{p}}$, and averaged over 20 repeated measurements. The bottom-most curve (black) is the background-subtracted Landau-level (LL) spectrum at $8$~T. The background for the 8.0-T curve is shown in violet. 
	In the background-subtracted LL spectrum, six peaks can be clearly identified and give the energy positions of the LLs. The LL peak marked with a blue circle can be tracked for different fields in the raw data shown in red. The peak position, which gives the eigen-energy $E_N$ of each LL, is determined by fitting a single Gaussian to each peak and the error in $E_N$ from the fit is always less than 0.2 meV. The combined fitting result is shown in thin grey line. (d) The eigen-energy $E_N$ of the six LLs identified in the STS data for all $B$-field values are plotted as a function of $\sqrt{NB}$, where $N$ is the LL index (top axis); the corresponding momentum $k$ on the TSS is shown on the bottom axis. A linear fit (red line) with Eq. (\ref{eq:LL_TSS}) gives $E_\mathrm{D}= -306$~mV and $v_\mathrm{F}= 5.8\times 10^{5}$~m/s.
	}
	\label{fig:STM_LL}
\end{figure}

To further validate our assignment of the spectral features in our \diduiu{} data, we have performed additional spectroscopic characterization of the surface electronic structure by mapping the spatial variations of the LDOS.
Typical \dIdU{} maps taken at the indicated bias voltages are shown in \figs{fig:STM_LL}(a) and (b). The spatial modulation of the LDOS due to quasiparticle interference (QPI), as opposed to structural effects, is evident due to the decrease of the wavelength of the QPI patterns as the bias voltage is increased.
Based on the band structure depicted in \fig{fig:STM_STS}(c), the QPI at the indicated bias voltages can be due to scattering of carriers in the BCB or TSS. While contributions of scattering bulk carriers can not be ruled out, we will show in the following that the dominant contribution is due to the TSS.

For the TSS, the largest possible scattering vector \textbf{q} is related to the wavevector \textbf{k} through \textbf{q}~$=2$\textbf{k}. However, the condition \textbf{q}~$=2$\textbf{k} corresponds to $180^\circ$ backscattering, which is strongly suppressed for a TSS with spin-momentum locking. Therefore, the dominant scattering vectors of the TSS will be smaller \cite{Zhang2009,Kuroda2010,Wang2011_qpi}. Even for \BiSe{}, it has been predicted that the hexagonal warping of the TSS~\cite{Fu2009} will open new scattering channels at energies sufficiently above the DP \cite{Kuroda2010}. Since the strength of the warping term in \SrBiSe{} is unknown, we simply use \textbf{q}~$\approx1.5$\textbf{k} (which is known for the more strongly warped TSS of $\mathrm{Bi}_2\mathrm{Te}_3$ \cite{Wang2011_qpi}) as a lower bound for the expected scattering vector length and \textbf{q}~$=2$\textbf{k} as the upper bound.
These two \textbf{q}-vectors are indicated by semicircles in the the Fourier transform (FT) of the \dIdU{} maps [insets of \figs{fig:STM_LL}(a) and \ref{fig:STM_LL}(b)] by taking $|\mathbf{k}|$ at the relevant energy from the TSS dispersion depicted in Fig. \ref{fig:STM_STS}(c).

We only observe clear QPI at bias voltages $\geq 100~\mathrm{mV}$, \textit{i.e.} more than 400~meV above the DP. At this energy the iso-energy surface of the TSS of \BiSe{} has a hexagonal shape \cite{Kuroda2010} and the TSS acquires a significant out-of-plane spin-polarization, enabling scattering vectors connecting opposite sides of the iso-energy surface mainly through the vertical spin component~\cite{Zhang2009,Fu2009}. For the hexagonal iso-energy surface the expected \textbf{q} vector is still close to $2$\textbf{k}, which is in agreement with the substantial intensity in the FT of \fig{fig:STM_LL}(a) near \textbf{q}~$=2$\textbf{k}. Extrapolating the iso-energy surface to 700~meV above the DP by considering the simple warping term \cite{Fu2009} extracted from the data for \BiSe{} \cite{Kuroda2010}, one expects a snowflake-like shape which is better known for $\mathrm{Bi}_2\mathrm{Te}_3$ \cite{Wang2011_qpi,Zhang2009,Fu2009}. The scattering vector \textbf{q}~$\approx1.5$\textbf{k} would dominate in such an iso-energy surface, which is in agreement with the FT of \fig{fig:STM_LL}(b). However, we note that at these high energies above the DP hybridization of the TSS with bulk bands certainly needs to be considered explicitly and will lead to modification of the dispersion relation of the TSS that go beyond what is captured by a simple warping term. Moreover, since at these energies the iso-energy surfaces not only have contributions from the TSS but also from the BCB~\cite{Kuroda2010}, scattering of bulk electrons and interband scattering between TSS and BCB may also contribute to the observed QPI.

Given these difficulties in interpreting the observed QPI, we have also examined the response of the LDOS to an external magnetic field of more than 8~T applied along the surface normal. Under these conditions, electrons are quantized into Landau levels. In the case of Dirac electrons of the TSS, the energy $E_N$ of the $N$th LL is to first approximation given as
\begin{align} 
	\label{eq:LL_TSS}
	E_{N} = E_{D} + \sgn{(N)}v_\mathrm{F}\sqrt{2eB\hbar \left| N \right|},	
\end{align} 
where $N$ is the Landau level index, $v_\mathrm{F}$ is the Fermi velocity, $B$ is the magnetic field and $E_\mathrm{D}$ is the Dirac point energy.

In \fig{fig:STM_LL}(c), a set of Landau levels is visible in the LDOS in a range of $-25$~meV to $+25$~meV around the Fermi energy. 
In order to extract the energy positions of the peaks, we subtracted the background from the STS curves using a cubic spline fit for the background (in violet), and each peak was fitted using a single Gaussian function. The background-subtracted data (black) are shown for $8.0$~T along with the raw data with fits (grey).
We identified the positions of six LLs and plotted their energy positions as a function of $\sqrt{NB}$ (in the range of 8.0 to 8.7 T) in \fig{fig:STM_LL}(d). By assigning the LL index of 28 to the highest peak, we can fit all the peak positions to Eq.~(\ref{eq:LL_TSS}) with reasonable values of $E_\mathrm{D}$ ($-306$~mV) and $v_\mathrm{F}$ ($5.8\times10^{5}$~m/s). The close-up of the fit near the data points is shown in the inset of \fig{fig:STM_LL}(d); although the data points are very linear in this plot of $E$ vs $\sqrt{NB}$, one cannot definitely tell from the data alone if the dependence on $N$ is linear (which is expected for BCB) or $\sqrt{N}$ (expected for TSS). Nevertheless, the constraint of the DP energy and $v_\mathrm{F}$ allows us to elucidate that the $\sqrt{N}$ dependence gives a more consistent analysis (see Appendix~\ref{sec:AppendixLL}). Therefore, we conclude that the observed LLs confirm that the spectral features at these energies are dominated by the TSS.

\subsection{Superconductivity of the surface}
\label{SC_tip}

\begin{figure}[t!]
	\includegraphics[width=8.6cm]{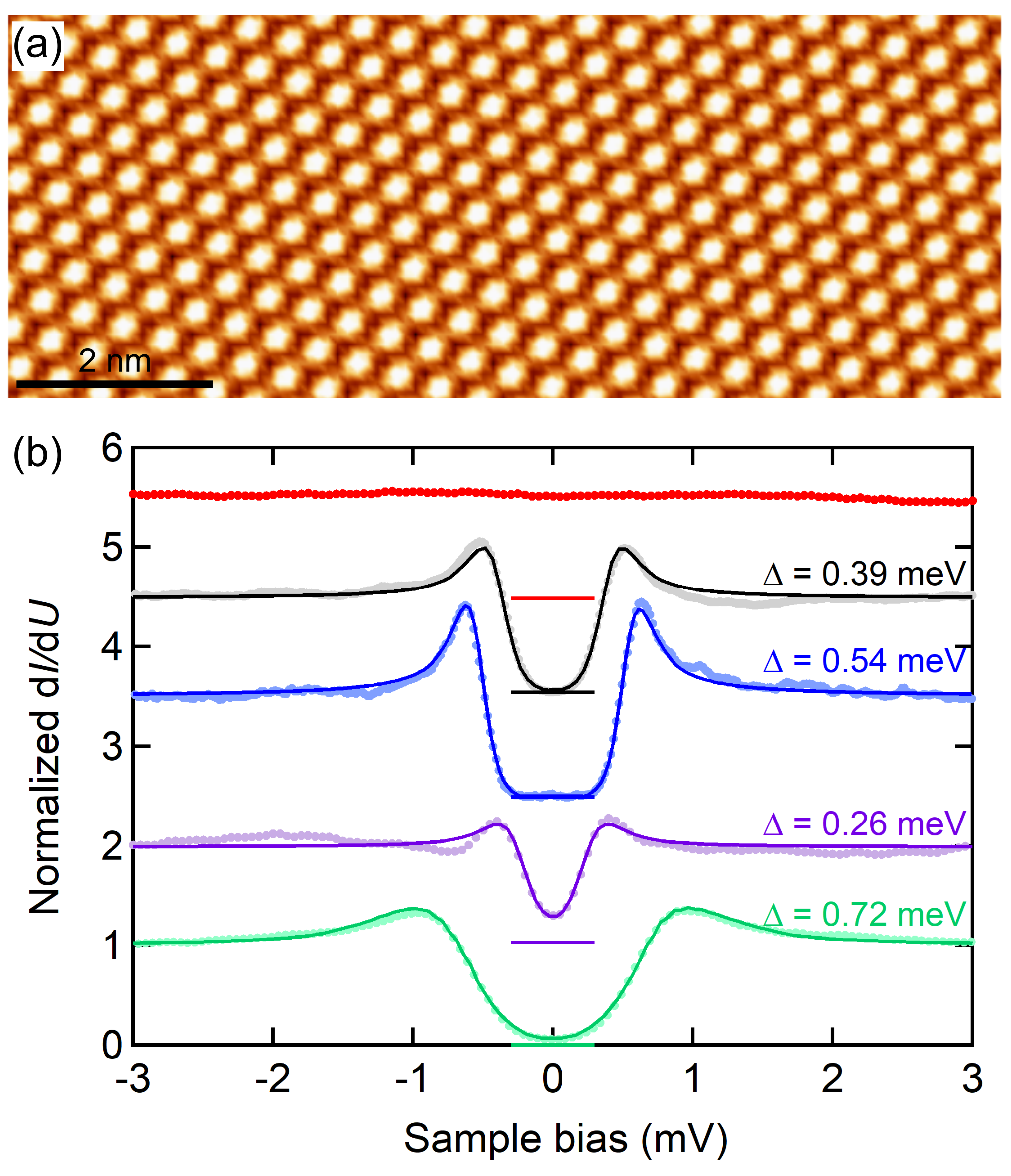}
	\caption{(a) Atomic-resolution image on the topmost Se layer. (b) Representative high resolution spectra (solid circles) taken with different tips at different positions on the surface: Substantial variations ranging from a flat LDOS at the Fermi level (red curve) to a full gap (blue curve) were observed. Spectra are offset vertically for clarity. Fitting of the data to the Dynes formula \cite{Dynes1978} yields the superconducting gap $\Delta$ of $0.39$~meV (black), $0.54$~meV (blue), $0.26$~meV (violet), and $0.72$~meV (green). Scan/stabilization parameters: (a)  $U=-900$~mV, $I=200$~pA; (b) $U=5$~mV (red, black, green, and violet), $U=3$~mV (blue), $I=200$~pA (red), $I=200$~pA (black), $I=500$~pA (blue), $I=100$~pA (violet), $I=25$~nA (green), $U_{\text{mod}}= 50$~$\mu$V$_{\text{p}}$. Except for the green curve taken at 1.7 K, all the spectra were taken at 0.4 K. The effective temperature $T_{\text{eff}}$ in the fit for the black, blue, and violet fit is 0.7~K, while that for the green curve is 2~K. The $\Gamma$ value for the fits are $0.02$, $0.0001$, $0.05$, and $0.0003$ meV for black, blue, violet, and green curves, respectively.
	}
	\label{fig:STM_gapsizes}
\end{figure}

With our understanding of the electronic structure at the surface of \SrBiSe{} established in the previous subsection, we now turn to the superconducting properties.
All spectroscopy experiments to this end were done at a nominal system temperature of about $0.4$~K and spectra were acquired on defect-free parts of the surface such as the one depicted in \fig{fig:STM_gapsizes}(a). 

Surprisingly, high resolution spectroscopy taken around the Fermi energy shows substantial variations.
Representative spectra are gathered in \fig{fig:STM_gapsizes}(b) where each trace corresponds to a spectrum taken with a different tip. 
Some spectra exhibit a flat LDOS [red trace in \fig{fig:STM_gapsizes}(b)] while others show a superconducting gap but with various gap sizes [black, blue, violet and green traces in \fig{fig:STM_gapsizes}(b)]. 
We have quantified the superconducting gap by fitting the spectra using the Dynes formula \cite{Dynes1978}. The differential conductance is given by:

\begin{align} 
	\label{eq:dynes1}
	G_{\text{N}}\frac{\partial}{\partial V} \int_{-\infty}^\infty N_{\text{S}}(E) \left[f(E,T_\mathrm{eff}) - f(E-eV,T_\mathrm{eff})\right]dE,	
\end{align} 
with $G_{\text{N}}$ the normal-state conductance, $f(E,T_\mathrm{eff})$ the Fermi function, and $N_{\text{S}}(E)$ the density of states in the BCS theory given as
\begin{align} 
	\label{eq:dynes2}
	N_{\text{S}}(E)= Re(\frac{(E-i\Gamma)}{\sqrt{(E-i\Gamma)^{2}-\Delta^{2}}}),	
\end{align} 
where $\Gamma$ is an effective broadening parameter and $\Delta$ is the superconducting gap.
The effective temperature $T_\mathrm{eff}$ is determined independently (see Appendix~\ref{app:AppendixTeff}).
We have used a total of 10 different new or freshly prepared PtIr tips and one W tip. 
Across all tips, the minimum and maximum $\Delta$ values observed were $0.19$ and $0.73$~meV, respectively. 
Our best fits yield $\Gamma$ values that are always below our energy resolution of about 100~$\mu$eV.

The large scatter in the superconducting gap size at the surface is unexpected for a sample with a sharp bulk superconducting transition. Indeed, the sharp transition observed in the bulk suggest homogeneity in the sample and therefore a more or less uniform gap size of $\Delta\approx1.76k_\mathrm{B}T_\mathrm{c}\approx0.4$~meV is expected.
As already mentioned in the introduction, a similar discrepancy between the superconducting gaps measured by STM and that expected from the superconducting transition temperature of the bulk has been reported in several publications on doped \BiSe{} compounds, a summary is shown in table~\ref{tab:sc_STM}.

\begin{figure}[t!]
	\includegraphics[width=8.6cm]{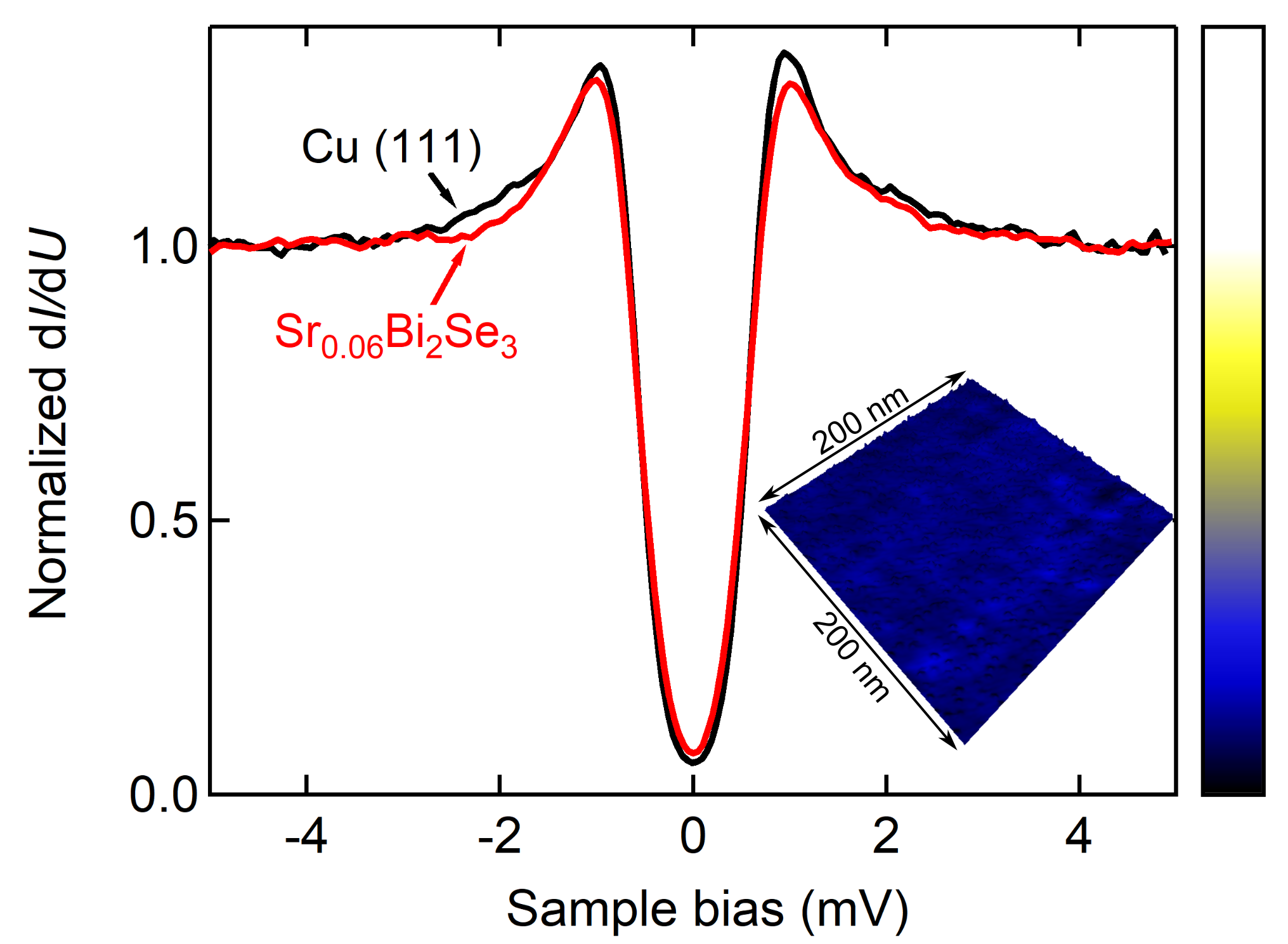}
	\caption{Normalized \dIdU{} spectrum taken with a PtIr tip on \SrBiSe{} showed a superconducting gap (red curve) after prolonged scanning. The same gap was observed on Cu(111) surface (black curve) demonstrating that the tip must be superconducting. Stabilization parameters are $U=5$~mV, $I=200$~pA (red), $I=25$~nA (black), $U_{\text{mod}}= 50$~$\mu$V$_{\text{p}}$.  The inset shows a topograph superimposed with a spectroscopy grid of 20 by 20 points, taken at a sample bias of 0~mV, at an applied magnetic field of 0.3~T. A homogeneous \dIdU{} signal close to zero is observed even in the presence of a magnetic field, which points to the absence of vortices and suggests that the superconducting gap originates not from the sample surface but from the probe tip.
	}
		\label{fig:STM_novortex}
\end{figure}

To make sure that superconducting gaps measured at the surface with STM are related to the superconducting state of the bulk, we applied an external magnetic field. 
Since \SrBiSe{} is known to be a type-II superconductor \cite{Shruti2015}, vortices are expected to be generated in applied fields greater than the lower critical field $\mu_0 H_{\text{c1}}$. In the inset of \fig{fig:STM_novortex} we show a $200$~nm by $200$~nm topography overlaid with a mapping of normalized differential conductance at zero bias taken in an applied field of 0.3~T. One can infer that the LDOS is completely uniform in this area,\textit{ i.e.} no vortex is observed in the entire field of view.

The absence of any vortex formation strongly suggests that the superconducting state probed by STM differs from the one in the bulk of \SrBiSe{}. Before addressing the origin of this difference, we first need to establish whether the superconducting gaps observed by STM are an intrinsic property of the surface or an artifact due to the probe tip.
For this purpose, we have first taken a high resolution spectrum on \SrBiSe{} (red trace in \fig{fig:STM_novortex}), and thereafter exchanged the sample against a non-superconducting copper sample and repeated the measurement with the same tip on the Cu(111) surface (black trace in \fig{fig:STM_novortex}). 
It is evident that the superconducting gap observed with this tip is essentially identical for both \SrBiSe{} and Cu(111) surfaces, thereby establishing unambiguously that the superconducting gap originates not from the sample surface but from the probe tip.

Having clarified that the superconducting gaps reproducibly observed on \SrBiSe{} are due to superconducting probe tips and \textit{not} due to superconductivity of the \SrBiSe{} surface itself, we now address the mechanism by which the tips made out of non-superconducting PtIr (or W) turn superconducting.

\begin{figure*}[]
	\centering
	\includegraphics[width=17.2cm]{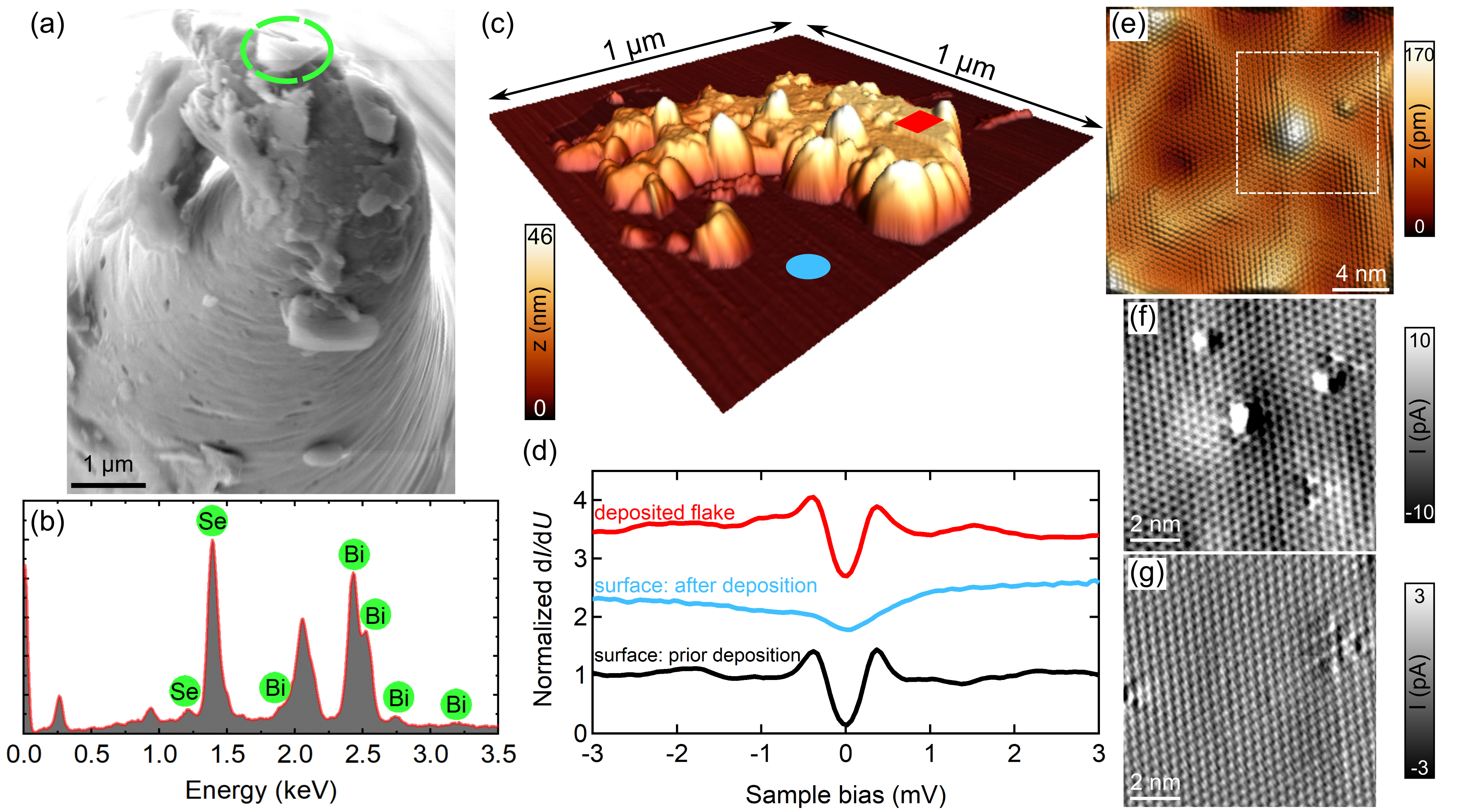}
	\caption{(a) SEM image of the apex of a PtIr tip after scanning on the sample. The apex is found to be covered with several micron-sized flakes. The dashed green circle marks the flake on which the EDX measurement shown in (b) was performed. (b) Result of the EDX analysis of the flake showing the Se peak at $1.379$~keV and the Bi double peaks at $2.423$~keV and $2.526$~keV. (c) 3D-rendered STM image of a flake deposited on the flat Se layer by a very mild collision between the tip and sample. The deposited material has an apparent height of $23$~nm and covers an area of roughly $0.2$~$\mu$m$^{2}$. The red box marks the area where the atomic-resolution image (e) of the flake was obtained. \scanp{} $U=-3$~V, $I=10$~pA for (c); $U=-900$~mV, $I=100$~pA for (e). The dashed white square in (e) encloses the area where the current image (f) was taken.
(d) Black curve shows the superconducting gap measured prior to the deposition of the flake; the area where the spectrum was taken [shown in the image (g)] presents unstrained Se lattice. After the deposition of the flake, the point STS (blue curve) taken on the area of the Se layer marked by blue circle in (c) shows only a weak proximity-induced gap, indicating that the tip LDOS is not gapped under the apex configuration after the flake has left. Nevertheless, the same tip apex measures a superconducting gap (red curve) on the deposited flake in the area marked by the red box in (c).  Stabilization parameter for (d): $U=5$~mV, $I=100$~pA, $U_{\text{mod}}= 50$~$\mu$V$_{\text{p}}$. Scan parameters for the current images: $U=-900$~mV, $I=100$~pA for (f); $U=5$~mV, $I=100$~pA for (g)}.
	\label{fig:STM_flake}
\end{figure*}

To this end, we have characterized a probe tip which showed a superconducting gap during STM experiments in more detail:
Figure \ref{fig:STM_flake}(a) shows an SEM image of the PtIr tip apex that is found to be covered with micron-sized flakes. The EDX spectrum [Fig.~\ref{fig:STM_flake}(b)] taken on a micro-flake marked by a dashed green circle shows prominent Bi and Se peaks, thus establishing that materials from the \SrBiSe{} crystal have been transferred onto the tip. Combined with the observation that a superconducting energy gap was measured with this probe tip prior to SEM characterization, one may conclude that the micron-sized \SrBiSe{} flakes found on the PtIr tip are superconducting.

To further support this conclusion, we have performed additional STM experiments in which we attempted to redeposit a flake onto the sample surface. Although no reproducible procedure could be established to this end, we found that mild tip-sample interactions sometimes lead to an accidental redeposition of a flake as shown in \fig{fig:STM_flake}(c). This flake has an apparent height of $\sim$25~nm and it extends by about one micron. 
High-resolution imaging on a flat area of the flake [red square in \fig{fig:STM_flake}(c)] clearly shows atomic resolution. The resolved hexagonal lattice locally has a lattice constant of about $0.4\, \mathrm{\AA}$ in agreement with the Se layer of \SrBiSe{}. However, one may also recognize nano-scale modifications of the surface height. We have thus compared the average in-plane atom densities of the flat part of the flake [\fig{fig:STM_flake}(f)] with the \SrBiSe{} surface prior to redeposition of the flake [\fig{fig:STM_flake}(g)]. Specifically, we have counted $613~\mathrm{atoms} / 100$~nm$^2$ in \fig{fig:STM_flake}(f) and $627~\mathrm{atoms} / 100$~nm$^2$ in \fig{fig:STM_flake}(g). This yields a reduction of about 2\%. Note that the observed difference in average in-plane packing density only demonstrates that there must be some strain in the flake and it does not mean a homogenous tensile strain of 2\%. Indeed, a detailed analysis of grain boundaries in \BiSe{} given in Ref.~\cite{Liu2014} showed variations in the magnitude of in-plane strain ranging from 20\% to $-20$\% occurring on nanometer length scales. Extracting similar quantitative values of the strain-tensor of the flake requires additional data and theoretical modeling, which is beyond the scope of this work.

The \dIdU{} spectra taken prior to the redeposition of the flake, an example of which is shown in black in \fig{fig:STM_flake}(d), clearly presents a superconducting gap; we observed similar spectra everywhere on the \SrBiSe{} surface. However, after the redeposition of the flake, the situation changed: While the spectrum shown in red in \fig{fig:STM_flake}(d), which was taken on the redeposited flake [in the area marked by the red square in \fig{fig:STM_flake}(c)], presents a superconducting gap similar to that observed prior to redeposition, the spectrum shown in blue, which was taken outside of the flake (in the region marked by the blue circle), does not present a fully-developed gap. 
These spectroscopic observations are consistent with the interpretation that a flake formed the tip-apex prior to redeposition and it is superconducting both before and after the redeposition.

\section{Discussion}

As already mentioned, the Hall resistivity data of our \SrBiSe{} crystal points to the electron density of $\sim$$4 \times 10^{19}~\mathrm{cm}^{-3}$ corresponding to the Fermi energy of $160$ to $270$~meV measured from the conduction band bottom, whereas the STS data on the same crystal shows that at the surface, the Fermi energy is only $\sim$100~meV from the conduction band bottom. This difference indicates unambiguously that there is an upward band bending of the BCB present at the surface. This upward band bending is partly due to the an intrinsic effect always present in \BiSe{} due to charge equilibration between bulk-like states and the TSS~\cite{Fregoso2015, Rakyta2015}, which can also be viewed as a result of many-body Coulomb interactions between the bulk and surface electrons~\cite{Wray2011}.
Note that band bending at the surface can also reflect additional factors~\cite{Zhan2012}, such as the contact potential due to the interface with the STM tip~\cite{Feenstra2006} or by adsorbates. In Cu-doped \BiSe{}, the Coulomb interactions between bulk and surface electrons was claimed to cause an upward band bending of the BCB by about 200 meV at the surface within the length scale of about 1 nm~\cite{Wray2011}.

The contact potential can be roughly estimated from the difference in work function of the metallic tip $\phi_m$ and that of \BiSe{}. Importantly, the latter is large $\phi_{\mathrm{Bi}_2\mathrm{Se}_3}\approx5.6$~eV~\cite{Takane2016} so that $\phi_{\mathrm{Bi}_2\mathrm{Se}_3}>\phi_{m}$ is generally fulfilled. Hence, the contact potential would only lead to the bulk bands bending down at the surface and it cannot be the cause of the observed upward band bending.

Intuitively, band bending due to adsorbates will lead to upward or downward band bending for adsorption of acceptor or donor molecules, respectively.
In this context, in particular photoemission experiments suffer from photoexcited adsorbate layers~\cite{Frantzeskakis2017}, that act as electron donors and, similar to the adsorption of alkali metals~\cite{Zhu2011}, inevitably cause downward band bending. However, the experimental conditions of our STM measurements ensure an adsorbate-free surface and more generally, downward band bending is incompatible with the experimental observations.

Importantly, while the details regarding the origin of the observed band bending can be complex, there is agreement that the experimentally measured surface potential of $\sim100$~meV will be screened over a typical length scale of $\sim 1$~nm  \cite{Fregoso2015, Rakyta2015,Wray2011}; in other words, near the surface, the bulk electrons experience an electric field of the order of $10^8$~V/m.

The puzzling fact is that the bulk of our \SrBiSe{} crystals show robust superconductivity with $T_\mathrm{c}$ = 2.8~K and a high shielding fraction \cite{Lin2021}, while the STS measurements found no superconducting gap anywhere on the surface down to a temperature of 0.4~K, when clean and non-superconducting probe tips are used. This absence of superconductivity on the surface has been a problem in many STM experiments performed on doped 
\BiSe{} superconductors, but the present study found that during the prolonged scanning of the tip on the \SrBiSe{} surface, the tip will always accumulate micro-flakes of \SrBiSe{} which show superconductivity. This finding helps to clarify some discrepancies in the observed SC gaps in earlier STM studies \cite{Kumar2021,Du2017,Han2015}, which were inconsistent with the bulk $T_\mathrm{c}$. Note that our experimental observation is essentially consistent with those on  superconducting Tl-doped Bi$_2$Te$_3$ and Nb-doped \BiSe{} reported by Wilfert \etal~\cite{Wilfert2018}, who concluded that superconductivity does not extend to the surface in these superconductors.

\begingroup
\begin{table*}	
	\setlength{\tabcolsep}{6pt}
	\renewcommand{\arraystretch}{1.2}
	\begin{threeparttable}
	\caption{Overview of surface studies on superconducting doped \BiSe{} materials.}
	\begin{tabular}{c c c c c c c}
		\hline\hline
		X$_{x}$Bi$_2$Se$_3$ & Tip & Bulk $T_\mathrm{c}$~(K) & $\Delta$~(meV) & Surface SC confirmed\footnote{Observation of a vortex lattice along with identification of NSC/SC boundary with the same tip apex.} & SC material on tip & TSS confirmed\\
	    \hline
	    Cu$_{\text{0.2}}$ \cite{Levy2013}* & Ir & 3.65 & 0.4 / 0.6\footnote{The gap of 0.6~meV was attributed to a superconducting tip.} & Yes / - &  No / Yes & No \\
	    Cu$_{\text{0.31}}$ \cite{Tao2018}* & PtIr & 3 & 0.46 / 0.77 & Yes / No & No / -  & No \\
	    Nb$_{\text{0.25}}$ \cite{Sirohi2018}* & - & 3.5 & 0.24 - 0.76\footnote{All SC gaps presented are taken into consideration.} & No & - & No \\
	    Nb$_{\text{x}}$ \cite{Wilfert2018}* & W & - & 0.79 & No & Yes & Yes \\
	    Nb$_{\text{0.25}}$ \cite{Qiu2015}** & - & 3.4 & - & - & - & Yes \\
	    Tl$_{0.06}$Bi$_2$Te$_3$\cite{Wilfert2018}* & W & 2.3 & 1 & No & Yes & Yes \\
			Sr$_{\text{0.08}}$ \cite{Han2015}** & - & 2.4 & 0.52 & No & - & Yes \\
	    Sr$_{\text{0.2}}$ \cite{Du2017}* & - & 3 & 0.42 - 1.15\footnotemark[3] & No & - & Yes \\	    
	    Sr$_{\text{0.1}}$ \cite{Kumar2021}* & - & 2.9 & 0.19 - 0.31\footnotemark[3] & No & - & No \\
	    	Sr$_{0.06}$*\footnote{this work.} & PtIr & 2.8 & 0.19 - 0.73 & No & Yes& Yes\\
	    Sr$_{0.06}$*\footnotemark[4] & W & 2.8 & 0.26 - 0.37 & No & Yes & Yes\\
		\hline\hline
		\end{tabular}
	\label{tab:sc_STM}
	\begin{tablenotes}
		\small
		\item Experimental technique: *STM, **STM and ARPES
	\end{tablenotes}
     \end{threeparttable}
\end{table*}
\endgroup

Since recent experimental \cite{Simoni2018} and theoretical \cite{Solinas2021} works found a suppression of superconductivity under a strong electric field of the order of $10^8$~V/m in conventional superconductors, we speculate that the local electric field associated with the experimentally observed band bending has a similar effect on superconductivity in \SrBiSe{}.
Indeed, since the upward band bending is a consequence of the existence of the TSS, one would expect the electric field to disappear (and the superconductivity to extend to the surface) when the TSS is destroyed. In this regard, there have been interesting reports that the strain on the surface can destroy the TSS 
in \BiSe{} \cite{Liu2014, Das2021}. Motivated by these observations, we speculate that during the transfer of the \SrBiSe{} flakes onto the tip, the flakes experience mechanical strain and the TSS is destroyed, allowing the superconductivity to extend to the surface of the flake. Although a detailed analysis of the strain in the flakes on our STM tips is beyond the scope of this work, we note that the atomic-resolution imaging on the flat parts of a flake [\fig{fig:STM_flake}(c)] shows dislocation features and a reduced atomic packing density, indicative of a strain in the flake.

Since the surface potential is an intrinsic effect linked inherently to the presence of the TSS in doped \BiSe{} and hence should be a general phenomena in the family of doped \BiSe{} superconductors, our speculation, that the appearance of superconductivity on the surface is related to the loss of TSS due to strain, gives a clue to understand earlier STM studies that are listed in Table~\ref{tab:sc_STM}.

It is prudent to mention that the pioneering STM work by Levy {\it et al.} \cite{Levy2013} gave persuasive evidence for superconductivity extending to the surface of Cu-doped \BiSe{} crystals (\textit{i.e.} appearance of vortices in an applied magnetic field and the observation of a domain boundary between superconducting and non-superconducting regions); nevertheless, the topographic images of the superconducting domains in Ref. \cite{Levy2013} showed many structural defects such as step bunching and grain boundaries that are consistent with strains in their samples. Moreover, no evidence was shown for the TSS to remain intact on these surfaces. 
Another STM study \cite{Tao2018} of Cu-doped \BiSe{} likewise showed experimental proof (vortex lattice in an applied magnetic field) that superconductivity can extend to the sample surface; however, less then 4\% of the studied surface area exhibited superconductivity and the remaining 96\% of the surface showed no superconducting gap. Interestingly, the topographic images in Ref. \cite{Tao2018} showed lots of structural defects, such as non-quintuple-layer step heights, in particular for the superconducting regions. 

Han {\it et al.} \cite{Han2015} used ARPES to demonstrate that the TSS of their \SrBiSe{} samples was intact and the STS data showed a superconducting gap, but their data are also fully compatible with a superconducting flake having been transferred to the tip.
The superconductivity observed on the surface of \SrBiSe{} by STM in the works by Kumar {\it et al.} \cite{Kumar2021} and Du {\it et al.} \cite{Du2017} showed values of $\Delta$, $T_\mathrm{c}$, and $\mu_0 H_\mathrm{c2}$ that are incompatible with the bulk values; this problem can be straightforwardly reconciled if a superconducting flake was present on the tip. Interestingly, Kumar {\it et al.} \cite{Kumar2021} also performed hard point-contact spectroscopy and found an increase in $T_\mathrm{c}$ with increasing pressure. It would be instructive to clarify if the strain due to the pressure from the tip leads to a destruction of the TSS.

\section{Conclusion}

While our \SrBiSe{} crystals present robust bulk superconductivity with $T_\mathrm{c}$ = 2.8~K, our STM measurements at 0.4~K with a fresh tip found no superconducting gap on the surface. This result is similar to many previous STM experiments on doped \BiSe{} superconductors. To understand this discrepancy, we propose that the upward band bending of 60 to 170~meV, which we elucidated at the surface, is playing a key role: Because recent DFT calculations found \cite{Fregoso2015,Rakyta2015} that this upward band bending is an inevitable consequence of the existence of the TSS and hence is intrinsic to the electron doped \BiSe{}-family of materials, we argue that the electric field suppresses the superconductivity at the surface in doped \BiSe{} superconductors. 
In this regard it was found both experimentally \cite{Simoni2018} and theoretically \cite{Solinas2021} that a strong electric field can kill superconductivity in conventional superconductors.
 
Intriguingly, after prolonged scanning on the \SrBiSe{} surface, the STS data taken with all probe tips eventually showed a superconducting gap whose origin can be assigned to the probe tip itself, and the \textit{ex-situ} SEM/EDX analysis of the tip establishes that micron-sized flakes of \SrBiSe{} are transferred onto the tip apex during the scanning of the surface. Furthermore, we were able to redeposit a \SrBiSe{} flake back onto the \SrBiSe{} surface inside the STM and confirmed that the flakes which were transferred onto the tip are indeed superconducting. Since recent works reported that the TSS can be destroyed by strain \cite{Liu2014, Das2021} and we actually observed lattice distortions on the redeposited flakes, we speculate that the strain in the flakes picked up by the tips destroys the TSS and allows the superconductivity to extend to the surface of the flake. This speculation that strain allows the superconductivity to extend to the surface can explain many of the puzzles in the past STM experiments \cite{Levy2013, Tao2018, Han2015, Kumar2021, Du2017, Wilfert2018}.

\acknowledgments{This project has received funding from the European Research Council (ERC) under the European Union's Horizon 2020 research and innovation programme (grant agreement No 741121) and was also funded by the DFG under CRC 1238 - 277146847 (Subprojects A04 and B06) as well as under Germany's Excellence Strategy - Cluster of Excellence Matter and Light for Quantum Computing (ML4Q) EXC 2004/1 - 390534769.}

\appendix
\section{Effective temperature}
\label{app:AppendixTeff}
\begin{figure}[t!]
	\centering
	\includegraphics[width=8.6cm]{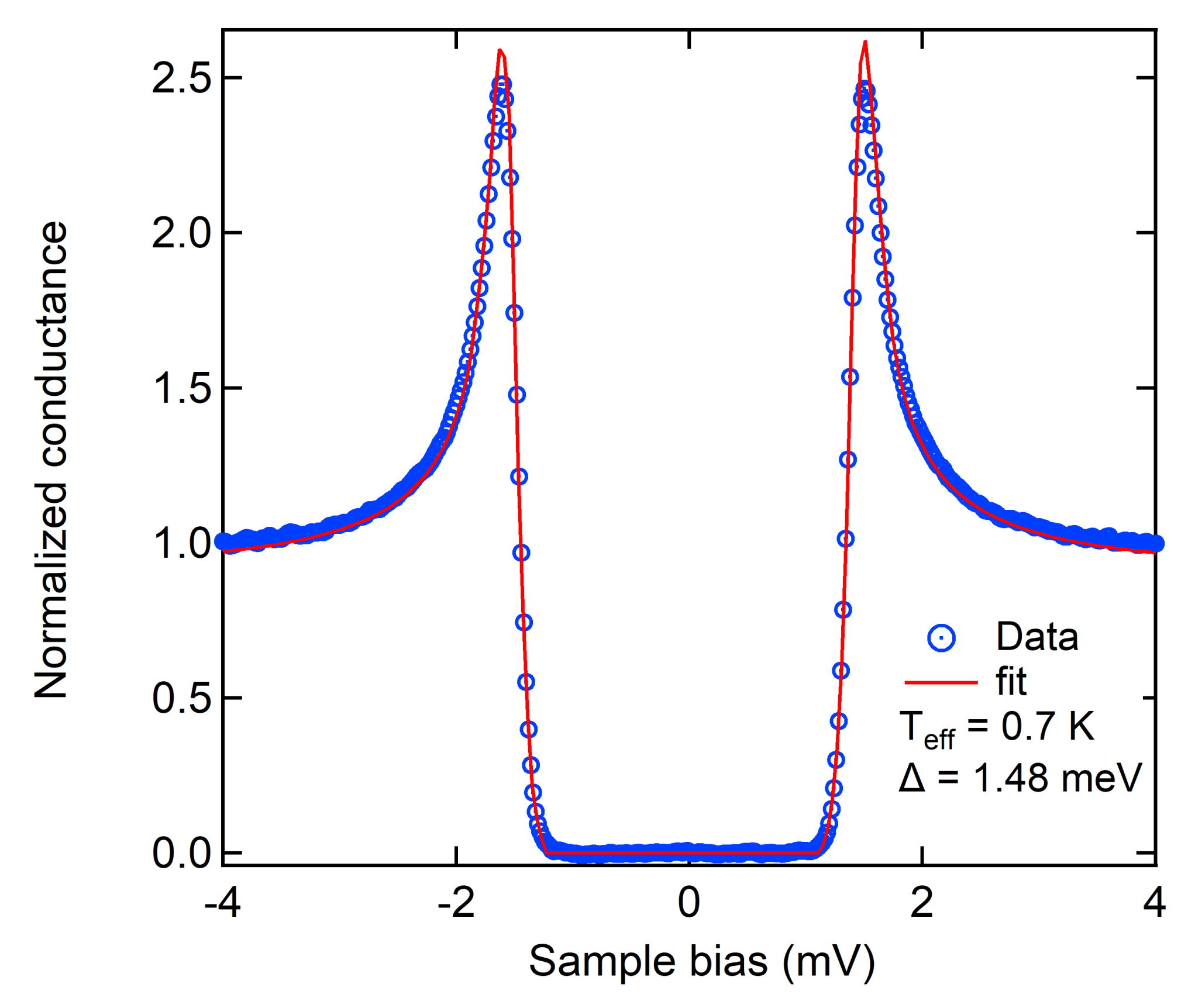}
	\caption{ STS taken with a Nb-tip on the Au(111) surface at $T=400$~mK. The superconducting gap of the Nb-tip (blue trace) and can be fitted using the BCS theory (red trace) with $\Delta=1.48$~meV which yields an effective electron temperature $T_{\text{eff}}=700$~mK. \stabp{}  $U=4$~mV, $I=500$~pA, $U_{\text{mod}}= 50$~$\mu$V$_{\text{p}}$.
	}
	\label{fig:STM_cal}
\end{figure}
At milli-Kelvin temperatures the nominal sample temperature differs from the effective electron temperature of the tunneling junction due to experimental broadening. This (generally unknown) experimental broadening has a similar effect in high resolution spectroscopy of SC gaps as the thermal broadening, hence one defines an effective temperature $T_{\text{eff}}$ which is calibrated by fitting the \dIdU\ spectrum of a well-known superconductor. For this purpose we fit the superconducting gap of a Nb-tip made of a high purity Nb wire measured on a non-superconducting Au(111) surface at a nominal temperature of 400~mK. We use the Dynes equation from the main text, but fix $\Gamma$ close to zero leaving only the effective temperature $T_{\text{eff}}$ and the superconducting gap $\Delta$ as fit parameters. 
The best fit to the experimental data is shown in \fig{fig:STM_cal}. The fit yields  $T_{\text{eff}}=0.7$~K.

\section{Analyses of the Landau level spectrum}
\label{sec:AppendixLL}

In the main text we argued that the peaks observed in the \dIdU{} data shown in \fig{fig:STM_LL} are due to Landau quantization of the TSS and not the BCB. Here, we compare the two scenarios in more detail. First, we recall that the reciprocal-space area $A(k)$ enclosed by a cyclotron orbit under the Landau quantization should satisfy the generalized Onsager relation
\begin{align} 
	\label{eq:Onesager}
	A_N(k) = 2\pi \frac{eB}{\hbar} (N + \lambda),
\end{align}
 where $N$ is an integer, $e$ is the elementary charge, $B$ is the magnetic field, and $\lambda=1/2-\gamma/(2\pi)$ with $\gamma$ the Berry phase.

For a circular orbit, Eq.~(\ref{eq:Onesager}) becomes
\begin{equation}
		\pi k_N^2 = 2\pi \frac{eB}{\hbar} (N + \lambda).
\end{equation}	
By using this $k_N^2$ in the parabolic dispersion relation $E=(\hbar k^2)/(2m_\mathrm{eff})$ for the BCB, the quantized energy levels for Schr\"odinger electrons are given by
\begin{equation}
	E_N = E_{\rm BCB} + \hbar \frac{eB}{m_\mathrm{eff}}\left(N+1/2-\frac{\gamma}{2\pi}\right),
\end{equation}
where $E_{\rm BCB}$ is the energy of the bottom of the BCB. 
Since $\gamma=0$ for Schr\"odinger electrons, one obtains
\begin{equation}
	E_N = E_{\rm BCB} + \hbar\omega_c(N+1/2).
	\label{eq:LL_BCB}
\end{equation}
On the other hand, for the linear dispersion relation $E = E_D + v_\mathrm{F}\hbar k$ of the Dirac electrons in the TSS, an analogous consideration yields
\begin{equation}
	E_N = E_D + v_\mathrm{F}\sqrt{2eB\hbar \left(|N|+1/2-\frac{\gamma}{2\pi}\right)}.
\end{equation}	
Since $\gamma=\pi$ for Dirac electrons, one obtains
\begin{equation}
	E_N= E_D + \sgn{(N)}v_\mathrm{F}\sqrt{2eB\hbar|N|},
\label{eq:LL_TSS1}
\end{equation}
which is already shown in the main text as Eq. (\ref{eq:LL_TSS}).

\begin{figure}[t]
	\centering
	\includegraphics[width=8.6cm]{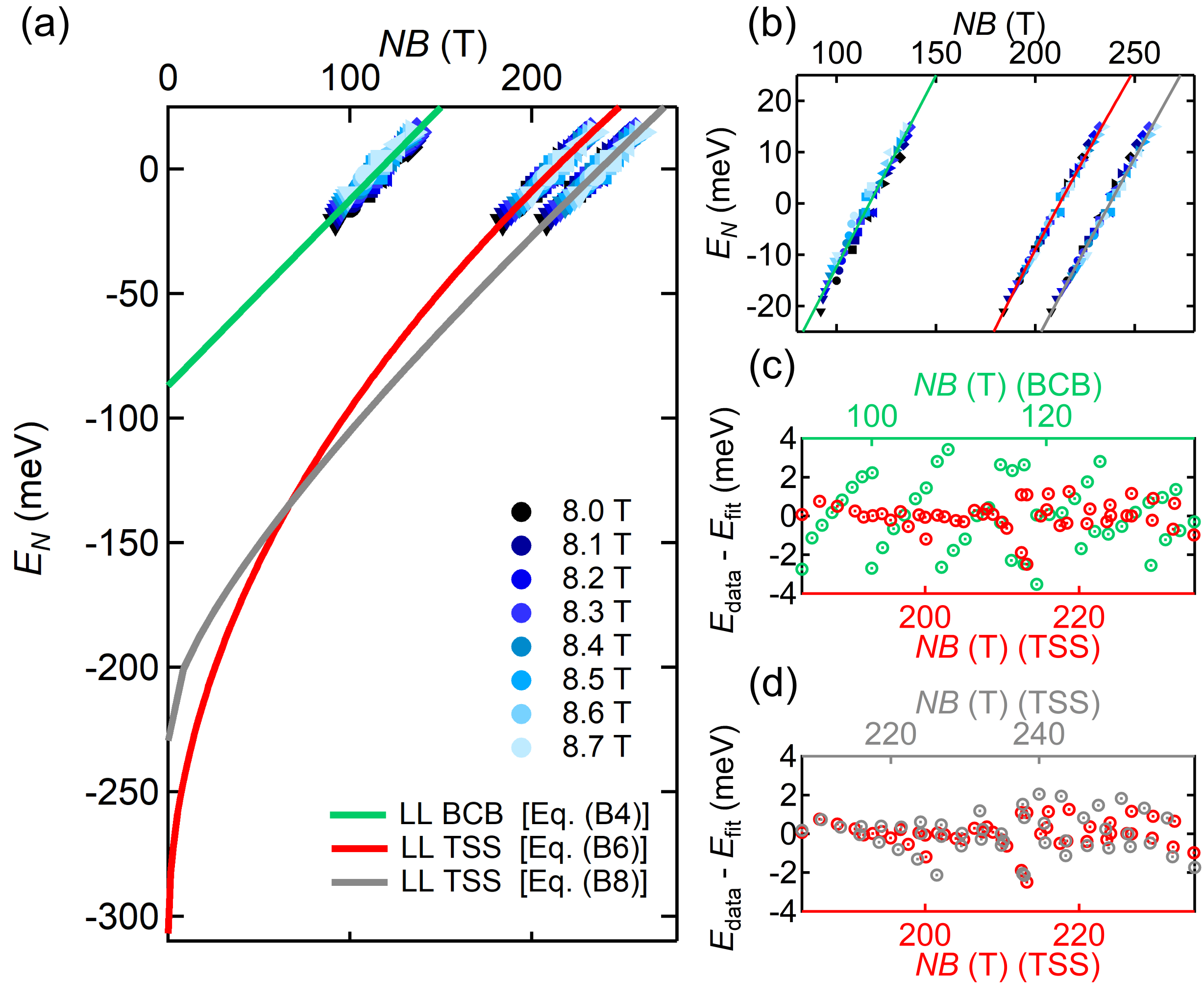}
	\caption{(a) The $E_N$ values of the six LLs identified in the STS spectra for all $B$ fields shown in Fig.~\ref{fig:STM_LL}(c) are plotted as a function of \textit{NB}, where $N$ is the LL index for the three cases under consideration (BCB, TSS with a linear dispersion, and TSS with a curved dispersion). A LL index of 16 is assigned to the highest peak for the case of the BCB, 28 for the case of a linear TSS dispersion, and 31 for the case of a curved TSS dispersion. The fit using Eq.~(\ref{eq:LL_TSS}) for the linear TSS dispersion is shown with the red curve and yields $E_\mathrm{D}= -306\pm3$~meV and $v_\mathrm{F}= 5.81 \pm 0.06 \times 10^{5}$~m/s. If the LL spectrum is assumed to originate from the BCB, the data should be fit using Eq.~(\ref{eq:LL_BCB}), and the best fit shown with the green linear line results in values of $-87\pm2$~meV for the bottom of the BCB and $0.155 \pm 0.004 m_e$ for the effective mass $m_{\mathrm{eff}}$. Consideration of a realistic TSS dispersion including a quadratic term leads to the fit using Eq.~(\ref{eq:LL_TSS_Taskin2}), and the best fit shown with the grey curve results in $E_\mathrm{D}= -230\pm5$~meV  and $v_\mathrm{F}= 2.13\pm 0.09\times 10^5$~m/s (assuming $m_{\mathrm{eff}}=0.25 m_e$ and $g =55$). A close-up of the three fits near $E_N$ = 0 meV is shown in (b). The deviations of the data from the fits are shown in (c) and (d), in which green symbols are for the BCB case, red symbols are for the linear TSS dispersion, and grey symbols are for the curved TSS dispersion.}
	\label{fig:TSS_BCB}
\end{figure}

Therefore, by extracting $E_N$ of the Landau levels, one can in principle distinguish between electrons stemming from the BCB (proportional to $N$) and those from the TSS (proportional to $\sqrt{|N|}$).
However, in the present experiment, the experimentally observed peaks in the LDOS due to Landau quantization are far above the band bottom of both BCB and TSS. Consequently, the relevant LLs located near the Fermi energy have relatively large indices. In such a case, the distinction between $\sqrt{N}$ [red curve in \fig{fig:TSS_BCB}(a)] and $N$ [green straight line in \fig{fig:TSS_BCB}(a)] becomes subtle due to the ambiguity in the assignment of $N$.

Nonetheless, we have made a trial to analyze the data shown in \fig{fig:STM_LL} assuming that they originate from the Landau quantization of the BCB whose bottom should be located about 100~meV below $E_\mathrm{F}$. 
The fit of the experimental data (assuming $N$ of the highest LL to be 16) to Eq.~(\ref{eq:LL_BCB}) shown with the green straight line in \fig{fig:TSS_BCB}(b) is considerably worse than for the analysis presented in the main text, which assumed the highest LL index of 28 and used Eq.~(\ref{eq:LL_TSS}), shown in  \fig{fig:TSS_BCB}(b) with the red curve.
We highlight the unsatisfactory agreement between the experimental data and the fit for the BCB scenario by plotting their deviations in \fig{fig:TSS_BCB}(c).

It is prudent to note that the dispersion of the actual TSS in \BiSe{} deviates from a simple linear function \cite{Taskin2011}. This deviation can be approximated by considering a quadratic term in the dispersion relation. Therefore, for completeness, we have also performed an analysis based on the dispersion relation

\begin{align}
E = E_D + v_\mathrm{F}\hbar k + \frac{\hbar^{2}}{2m_\mathrm{eff}}k^{2} .   \label{eq:LL_TSS_Taskin}
\end{align}

Using this dispersion and including the Zeeman energy, the following expression for the eigen-energies of the LLs is obtained for the electron branch of the curved Dirac cone \cite{Taskin2011}:

\begin{align}
	  E_N= E_D + \hbar\omega_cN + \sqrt{2\hbar v_\mathrm{F}^2 eBN + \left(\frac{\hbar\omega_cN}{2}  - \frac{g_{s}\mu_{B}B}{2} \right)^2 }.  \label{eq:LL_TSS_Taskin2}
	\end{align}
In the case of \BiSe{}, the $g$-factor of $g_{s} = 55$, the effective mass of $m_\mathrm{eff} = 0.25 m_e$ and the Fermi velocity of $v_\mathrm{F}=3\times10^5$~m/s are established to accurately describe Shubnikov-de Haas oscillations \cite{Taskin2011}.
We found that the LL spectrum observed in \SrBiSe{} can be well explained [grey curve in \fig{fig:TSS_BCB}(a)] by assuming a reduced Fermi velocity of $v_\mathrm{F}= 2.13\times 10^5$~m/s and the highest LL index of 31. 
The reduction in the Fermi velocity is in line with the case of superconducting Cu$_x$\BiSe{}, where a reduction of up to $30\%$ was observed \cite{Wray2011}. 

It should be remarked that in this analysis based on Eq. (\ref{eq:LL_TSS_Taskin}), we have to assume a Dirac-point energy of $E_\mathrm{D}= -230$~meV. While this is in apparent disagreement with the DP energy of about $-310$~meV deduced from the minimum in the LDOS shown in \fig{fig:STM_STS}, it can be straightforwardly reconciled by considering a tip-induced band bending that causes a shift in the DP energy; namely, as the bias voltage is increased, the electric field between the tip and the surface becomes stronger, causing the DP to shift in energy. Such a shift of about $-80$~meV was deduced previously for \BiSe{} by comparing STS and ARPES measurements~\cite{Cheng2010}. 
Hence, our experimental LL spectrum is well described by both models of the TSS, i.e. Eq. (\ref{eq:LL_TSS}) and Eq. (\ref{eq:LL_TSS_Taskin2}). Due to the ambiguity in the assignment of the LL index, a more rigorous distinction is difficult. Nevertheless, regardless of the model for the TSS, the observed LL spectrum supports the existence of upward band bending near the surface.

\end{document}